\documentclass[english,nofootinbib,notitlepage,superscriptaddress,10pt,aps,pra,twocolumn]{revtex4-1}
\usepackage[utf8x]{inputenc}
\usepackage{amsmath}
\usepackage{amssymb}
\usepackage{bbm}
\usepackage{graphicx}
\usepackage{babel}
\usepackage{color}
\usepackage{subfigure}

\begin{document}

\title{Dual redshift on Planck-scale-curved momentum spaces}

\author{Giovanni AMELINO-CAMELIA}

%\email{Giovanni.Amelino-Camelia@roma1.infn.it}

\affiliation{Dipartimento di Fisica, Universit\`a di Roma ``La Sapienza", P.le A. Moro 2, 00185 Roma, Italy}
\affiliation{INFN, Sez.~Roma1, P.le A. Moro 2, 00185 Roma, Italy}

\author{Leonardo BARCAROLI}

\affiliation{Dipartimento di Fisica, Universit\`a di Roma ``La Sapienza", P.le A. Moro 2, 00185 Roma, Italy}
\affiliation{INFN, Sez.~Roma1, P.le A. Moro 2, 00185 Roma, Italy}

\author{Giulia GUBITOSI}

\affiliation{Dipartimento di Fisica, Universit\`a di Roma ``La Sapienza", P.le A. Moro 2, 00185 Roma, Italy}
\affiliation{INFN, Sez.~Roma1, P.le A. Moro 2, 00185 Roma, Italy}

\author{Niccol\'o LORET}

\affiliation{Dipartimento di Fisica, Universit\`a di Roma ``La Sapienza", P.le A. Moro 2, 00185 Roma, Italy}
\affiliation{INFN, Sez.~Roma1, P.le A. Moro 2, 00185 Roma, Italy}

\begin{abstract}
Several approaches to the investigation of the quantum-gravity problem have provided ``theoretical evidence" of a role for the Planck scale in characterizing the geometry of momentum space. One of the main obstructions for a full exploitation of this scenario is the understanding of the role of the Planck-scale-curved geometry of momentum space in the correlations between emission and detection times, the ``travel times"
for a particle to go from a given emitter to a given detector. These travel times appear to receive Planck-scale corrections for which no standard interpretation is applicable, and the associated implications for spacetime locality gave rise to the notion of ``relative locality" which is still in the early stages of investigation. We here show that these Planck-scale corrections to travel times can be described as ``dual redshift" (or ``lateshift"): they are manifestations of momentum-space curvature of the same type already known for ordinary redshift produced by spacetime curvature. In turn we can identify the novel notion of ``relative momentum-space locality" as a known but under-appreciated feature associated to ordinary redshift produced by spacetime curvature, and this can be described in complete analogy with the relative spacetime locality that became of interest in the recent quantum-gravity literature.
We also briefly comment on how these findings may be relevant for an approach to the quantum-gravity problem proposed by Max Born in 1938 and centered on Born duality.
\end{abstract}

\maketitle

$~$

\newpage

$~$

\newpage

\section{Introduction}

Over the last decade several independent arguments pointed more or less explicitly toward a role for the Planck scale in characterizing a non-trivial geometry of momentum space (see, {\it e.g.}, Refs.~\cite{majidCURVATURE,dsr1Edsr2,jurekDSMOMENTUM,girelliCURVATURE,schullerCURVATURE,changMINIC,principle,grf2nd}).
Remarkably Max Born had already argued in  1938~\cite{born1938}, inspired by Born duality, that curvature of momentum space might be a needed step toward quantum gravity, but for several decades this had met little interest (see, however, Ref.~\cite{golfand}). Attention to this scenario was imposed by some of the most modern formalisms in use for the study of the quantum-gravity problem. For example, in the study of noncommutative spacetimes, particularly when considering models with ``Lie-algebra spacetime
noncommutativity", $[x_\mu , x_\nu]= i \zeta^\sigma_{\mu \nu} x_\sigma$, the momentum space on which spacetime coordinates generate translations is evidently curved (see, {\it e.g.}, Ref~\cite{gacmaj}). And also in the study of the Loop Quantum Gravity approach~\cite{rovelliLRR} one can adopt a perspective involving momentum-space curvature (see, {\it e.g.}, Ref~\cite{leeCURVEDMOMENTUM}).

In light of these results, it could be important for quantum-gravity research to identify the main observable consequences of momentum-space curvature. And indeed there has been a strong effort recently concerning what appears to be the best  candidate manifestation of Planck-scale momentum-space curvature: these are
studies~\cite{gacmaj,grbgac,ellisDELAY,leelaurentGRB,AntoninoGiulia,jackdesitter,kowajackcurvo} of the correlations between emission and detection times, the ``travel times" between a given emitter and a given detector. It is found that the Planck-scale curvature of momentum space introduces corrections to the travel times, opening also an opportunity for experimental tests~\cite{gacmaj,grbgac,ellisDELAY,leelaurentGRB,AntoninoGiulia,jackdesitter}. But the conceptual status of these Planck-scale corrections to travel times remains only poorly understood, and in particular it is emerging that a crucial role should be played by a correspondingly weaker notion of spacetime locality, the ``relative locality" of Refs.~\cite{principle,grf2nd,bob,leeINERTIALlimit,arzkowaRelLoc,kappabob,transverse}
whose understanding is just in the early stages of development.

We here show that these gray areas of our understanding of corrections to travel times due to Planck-scale curvature of momentum space and the associated relativity of spacetime locality can be  clarified by fully embracing the spirit of Born's 1938 proposal. These aspects of the implications of curvature of momentum space must admit a description which is just dual to known properties of theories with spacetime curvature.
And indeed we find that the travel-time features noticed in studies with momentum-space curvature
 admit description as ``dual redshift", which we shall here also label as ``lateshift": they are manifestations of momentum-space curvature of the same type already known for ordinary redshift produced by spacetime curvature.

\newpage

 In order to keep our presentation clear and compact we focus on the case of only 1+1 spacetime dimensions ($2D$), investigating the duality between 2D de Sitter spacetime (dS spacetime) and 2D de Sitter momentum space (dS momentum space).

 We derive very explicitly a description of the corrections to travel times due
  to the Planck-scale curvature of momentum space given in terms
  of  ``dual redshift". And also for relative spacetime locality produced by momentum-space curvature we find that it is
dual to a known but under-appreciated feature associated to ordinary redshift produced by spacetime curvature, which we here label ``relative momentum-space locality". The presence of relative spacetime locality
for theories with curved momentum space has been occasionally perceived with an aura of mystique and/or suspicion
 (see, {\it e.g.}, Ref.\cite{sabinePRL,sabinePERCACCI}), but our analysis clarifies that relative locality is
 a simple
 and intelligible consequence of cases in which the observers cannot or anyway do not adopt spacetime coordinates
that are conjugate to the generators of spacetime translations\footnote{We here focus on spacetime translations, but the careful reader will appreciate that the argument generalizes to any case in which the coordinates have nontrivial properties under the transformations of interest. For example, the analysis in Ref.\cite{bob} considers a curved momentum space and a pair of distantly boosted observers (connected by a boost in combination with a translation), and there the relative locality arises because the spacetime coordinates do not have simple properties under the action of compositions of boosts and translations.}.
And in fact we can point the attention of our readers to a well-known dual feature, which we feel
 deserves to be labeled as ``relative momentum-space locality",
which is present in the classic analyses of de Sitter spacetime and reflects indeed the fact that the curvature of de Sitter
spacetime can encourage the adoption of coordinates on momentum space that are not conjugate to spacetime coordinates.

\section{From redshift to lateshift}

We start by essentially summarizing our key results for lateshift, sketching out the duality that emerges from our analysis between the lateshift produced by momentum-space curvature and ordinary redshift produced by spacetime curvature.
Later sections will provide further details.
We illustrate the duality by using comoving coordinates on the 2D-dS-spacetime side and for the 2D-dS momentum space we use coordinates dual to those (``comoving on momentum space").
 The duality is centered on exchanging the expansion rate $H$ of dS spacetime for the inverse of the Planck scale, here denoted by $\ell$, which plays indeed the formal role of expansion rate on the dS momentum-space side of the duality.

\newpage

\subsection{Redshift}

\label{sec:redshift}

For dS  $1+1$-dimensional spacetime the metric takes the form ($\mu,\nu=0,1$)
\begin{equation}
ds^{2}=(dx^{0})^{2}-e^{2Hx^{0}}(dx^{1})^{2} \label{eq:dSmetric}
\end{equation}
\noindent while energy $p_0$ and spatial momentum $p_1$ are conjugate to the spacetime coordinates:
\begin{equation}
\{x^0,x^1\}=0, \;\; \{p_0,p_1\}=0, \;\; \{p_\mu,x^\nu\}=\delta_\mu^\nu\, .
\label{phasespaceONE}
\end{equation}
Spatial momentum is a conserved charge, for which we shall use equivalently the notation $p_1$ and $\Pi_1$. Energy is not conserved, because of spacetime expansion. Time translations are deformed by spacetime expansion, and the associated charge $\Pi_0$ has the properties
\begin{eqnarray}
\!\!\!\!\! \{\Pi_0,\Pi_1\}=H \Pi_1 \,,~~
\{\Pi_0,x^1\}=-H x^1 \,,~~
\{\Pi_0,x^0\}=1 ~.~~~~
\label{phasespaceTWO}
\end{eqnarray}
It is useful to notice that $\Pi_0 = p_0-Hx^1p_1$. And it shall be relevant for rendering more vivid our duality to observe that in dS spacetime (with comoving coordinates) the worldlines of massless particles crossing the origin of the observer take the form
\begin{equation}
x^1 = \frac{1-e^{-Hx^0}}{H}
\label{dS-wl}
\end{equation}
For a particle on such a worldline one has that energy and momentum are related through the particle's time coordinate:
\begin{equation}
p_{1}=- e^{H x^{0}}\, p_{0}
\label{effectiveonshellness}
\end{equation}
The conceptual content of redshift in dS spacetime is particularly intuitive when comparing results for energy measurements by two observers, say Alice and Bob, whose origins are connected by a worldline of type (\ref{dS-wl}).
 %We have two such observers if Bob is obtained from Alice by a pure translation of parameters $a^{0},a^{1}$ satisfying the relation %
%\begin{equation}
%a^{1}=a^{0}
%\label{a0a1DS}
%\end{equation}
Indeed redshift is an effect such that a blue particle emitted at some source reaches a distant telescope
as a red particle.
For easier comparison with the results we shall later derive for a curved momentum space,
we prefer to characterize quantitatively
the redshift effect due to spacetime curvature
 by comparing its effects on two different particles
emitted with the same energy (``both blue") and from the same source but at different times.
We therefore consider two particles emitted with the same
energy by emitter Alice (the two worldlines both cross Alice's {\underline{spatial}} origin)
and derive in Sec.~\ref{sec:deSitterDetails} the difference in energy of detection of these two particles
at some distant detector Bob (Bob is such that the two particles both cross Bob's spatial origin).
We find the following result:
\begin{equation}
\left\{
\begin{array}{l}
\tilde{p}_{0}^{@A}=p_{0}^{@A}\,,~~\tilde{x}^{0}_{@A} \neq x^{0}_{@A} \label{redshiftmaster}\\
\\
  \tilde{p}_{0}^{@B}=e^{-H [\tilde{x}^{0}_{@B}-x^{0}_{@B}]}p_{0}^{@B}\end{array}
\right.
\end{equation}
where $p_{0}^{@B}$ and $\tilde{p}_{0}^{@B}$ (respectively $x^{0}_{@B}$ and $\tilde{x}^{0}_{@B}$) are the energies (respectively the times) of detection at Bob, indeed for two particles
emitted at Alice with the same energy ($\tilde{p}_{0}^{@A}=p_{0}^{@A}$) but at different times ($\tilde{x}^{0}_{@A} \neq x^{0}_{@A}$).

%\begin{equation}
%\!\!\!\!\!\!\!\!\!\!\! \tilde{p}_{0}^{@A}=p_{0}^{@A}\,,~~\tilde{x}^{0}_{@A} \neq x^{0}_{@A}
% \Longrightarrow \tilde{p}_{0}^{@B}=e^{-H [\tilde{x}^{0}_{@B}-x^{0}_{@B}]}p_{0}^{@B}
% \label{redshiftmaster}
%\end{equation}

\subsection{Lateshift}\label{sec:lateshift}

In establishing the duality with the results in the previous subsection we of course describe the metric of ($1+1$-dimensional) dS momentum space as follows:
\begin{equation}
dk^{2}=(dp_{0})^{2}-e^{2\ell p_{0}}(dp_{1})^{2}
\label{metricDSMS}
\end{equation}
And we introduce spacetime coordinates as conjugate to the momenta:
\begin{equation}
\{x^1,x^0 \}=0, \;\; \{p_1,p_0\}=0,\;\;
\{x^\mu,p_\nu\}=\delta^\mu_\nu \,,
\label{phasespaceONEb}
\end{equation}
\noindent We shall keep the analogy as close as possible by also introducing (in analogy with $\Pi_0,\Pi_1$ of the previous subsection) some ``relative-locality coordinates" $\chi^{0},\chi^{1}$ with $\chi^{1} \equiv x^1$ and $\chi^{0}\equiv x^0 - \ell x^1 p_1$, so that
\begin{eqnarray}
 \! \{\chi^0,\chi^1\}=\ell \chi^1 \,,~~
\{\chi^0,p_1\}=-\ell p_1 \,,~~
\{\chi^0,p_0\}=1 \,.~~~ \label{dssimpl}
\end{eqnarray}
$\chi^{0}$ and $\chi^{1}$ generate the translational symmetries of the dS momentum space, but (again in analogy with
the dS spacetime case)  $\chi^{0}$ does not generate pure $p_0$ shifts.

We shall show that the on-shell condition for massless particles on the dS momentum-space takes the
form
\begin{equation}
p_1 = \frac{1-e^{-\ell p_0}}{\ell}\, ,
\label{onshellness}
\end{equation}
which is interestingly dual to the Eq.~(\ref{dS-wl}) for the worldlines of massless particles in dS spacetime,
while the dS-momentum-space picture of worldlines of massless particles is given by
\begin{equation}
x^1=-e^{\ell p_0} x^0
\label{kM-wl}
\end{equation}
which is interestingly dual to the Eq.(\ref{effectiveonshellness})
playing the role of on-shell relation on the dS-spacetime side.

These Eqs.(\ref{metricDSMS}),(\ref{phasespaceONEb}),(\ref{dssimpl}),(\ref{onshellness}),(\ref{kM-wl}) are exactly dual to the Eqs.(\ref{eq:dSmetric}),(\ref{phasespaceONE}),(\ref{phasespaceTWO}),(\ref{dS-wl}),(\ref{effectiveonshellness})
valid on the dS spacetime. This will prove sufficient for our purposes even though there is an element of our analysis that is not properly dual: properties of spacetime translations are responsible for both redshift and lateshift. An even more precise duality would be found if one studied the
implications of momentum-space curvature for momentum-space translations, but those are of limited interest in physics.
The duality between redshift and lateshift is nonetheless strong enough to allow us to derive in Sec.\ref{sec:kPoincareDetails}
a result, which we propose as main characterization of lateshift,
which is indeed dual to the characterization of redshift we gave in Eq.~(\ref{redshiftmaster}).
This is found by considering again an emitter Alice and a detector Bob, and takes the shape of the relationship
%We  find that the origin of an observer Bob is connected by the worldline (\ref{kM-wl}) to the origin of an observer Alice if Bob is %obtained from Alice by spacetime translation parameters $a^{0},a^{1}$ such that
%\begin{equation}
%a^{1}=-e^{\ell p_{0}}a^{0} \label{eq:k-translationParameters}
%\end{equation}
\begin{equation}
\left\{
\begin{array}{l}
 \tilde{p}_{0}^{@A} \! \neq \! p_{0}^{@A}\,,~\tilde{x}^{0}_{@A} \! = \! x^{0}_{@A} \label{lateshiftmaster}\\
 \\
 \tilde{x}^{0}_{@B} -\tilde{x}^{0}_{@A} \! = \! e^{-\ell [\tilde{p}_{0}^{@B} - p_{0}^{@B}]} (x^{0}_{@B} - {x}^{0}_{@A}) \, ,
\end{array}
\right.
\end{equation}
where, consistently with the duality we are exposing, for this result (\ref{lateshiftmaster})
we consider two particles emitted at the same time at Alice ($\tilde{x}^{0}_{@A} \! = \! x^{0}_{@A}$)
with different energies ($\tilde{p}_{0}^{@A} \! \neq \! p_{0}^{@A}$).

This result (\ref{lateshiftmaster}) evidently characterizes lateshift as the source of the peculiarities for
the correlations between emission times and detection times previously found in the curved-momentum-space
literature: Eq.~(\ref{lateshiftmaster}) confirms that in presence of momentum-space curvature
 two massless particles emitted simultaneously
 at Alice
with different energies ($\tilde{p}_{0}^{@A} \! \neq \! p_{0}^{@A}$) reach a distant detector Bob at different
times ($\tilde{x}^{0}_{@B} \! \neq \! x^{0}_{@B}$), indeed governed by (\ref{lateshiftmaster}).

We stress again that the relativistic duality between the two cases is exact: formulas (\ref{eq:dSmetric}),(\ref{phasespaceONE}),(\ref{phasespaceTWO}),(\ref{dS-wl}),(\ref{effectiveonshellness})
familiar for dS spacetime get mapped into the exactly dual formulas
(\ref{metricDSMS}),(\ref{phasespaceONEb}),(\ref{dssimpl}),(\ref{onshellness}),(\ref{kM-wl})
for the novel case of dS momentum space.
But the questions we typically ask experimentally to these exactly dual pictures are not exactly dual to each other: in both cases one is interested
in spacetime translations, since in both cases one is primarily considering situations with spatially distant emitter and detector.
The relativistic duality we are exposing is however so strong that it still affects very significantly
 the final results
 (\ref{redshiftmaster}) and (\ref{lateshiftmaster}).

It is also useful to observe
that the duality we are analyzing becomes trivial when no curvature is present: relativistic theories of Minkowski spacetime and relativistic theories of Minkowski momentum space coincide (our duality turns into a self-duality when curvature is absent).
In the Minkowski case  massless particles with any difference in energy $\tilde{p}_{0}^{@A} - p_{0}^{@A}$ emitted with any
emission-time difference $x^{0}_{@A} - \tilde{x}^{0}_{@A}$ at emitter Alice are then detected at some distant detector Bob (at rest with respect to the emitter) with same difference of detection times $x^{0}_{@B} - \tilde{x}^{0}_{@B}=x^{0}_{@A} - \tilde{x}^{0}_{@A}$
and the same difference of energies $\tilde{p}_{0}^{@B} - p_{0}^{@B} = \tilde{p}_{0}^{@A} - p_{0}^{@A}$.

\begin{figure}[h!]

\centering

\fbox{\includegraphics[scale=1]{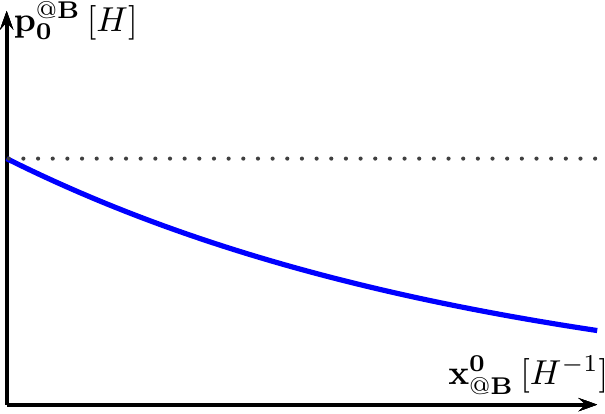}}
\caption{\footnotesize We here show how,
for fixed time and
energy of emission at Alice, there is a correlation
between the time of detection at Bob and the energy of detection at Bob, for the case of Minkowskian spacetime (dotted line)
and the case of dS spacetime (solid line).
The behaviour here shown for the dS-spacetime case, which is a characteristic manifestation of redshift,
is governed by Eq.(\ref{redshiftinfig}) here derived in the later Sec.\ref{sec:deSitterDetails}.}
\label{fig:Redshift_w_SR}
\end{figure}

\begin{figure}[h!]
\centering
\fbox{\includegraphics[scale=1]{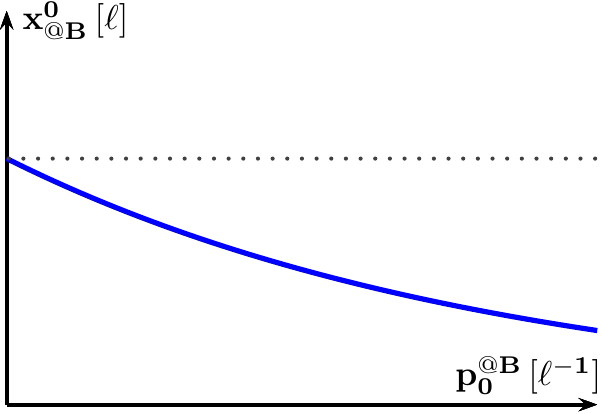}}
\caption{\footnotesize We here show how,
for fixed time and energy of emission at Alice, there is a correlation
between the time of detection at Bob and the energy of detection at Bob, for the case of Minkowskian momentum space (dotted line) and the case of dS momentum space (solid line).
The behaviour here shown for the dS-momentum-space case, which is a characteristic manifestation of lateshift,
is governed by Eq. (\ref{lateshiftinfig}) here derived in the later Sec.\ref{sec:kPoincareDetails}.}
\label{fig:Lateshift_w_SR}
\end{figure}

In Figs.\ref{fig:Redshift_w_SR} and \ref{fig:Lateshift_w_SR}
we visualize an aspect of the duality here exposed for the case when curvature is present, also in reference to the
 self-duality present when there is no curvature.
 For these visualization purposes we find useful to rely on the correlations that the analysis we present in later sections
 finds between time of detection at Bob and energy of detection at Bob of a massless particle. The presence of a horizontal dotted line in both
Fig.\ref{fig:Redshift_w_SR} and Fig.\ref{fig:Lateshift_w_SR} reflects the fact that in absence of curvature the energy of detection
is independent of the time of detection (in absence of curvature there is no redshift, so the energy of detection is automatically
the same as the energy of emission of the particle) and the time of detection is independent of the energy of detection
(in absence of curvature there is no lateshift, so the time of detection is given, for massless particles of any energy, in terms of the
distance between emitter and detector).

Fig.\ref{fig:Redshift_w_SR} also shows (solid line) the quantitative behaviour of redshift produced by spacetime curvature: for fixed time and
energy of emission at Alice there is a correlation (governed by the distance between Alice and Bob, left implicit in figure)
between the time of detection at Bob and the energy of detection at Bob.
This correlation is of course governed by the distance between Alice and Bob, left implicit in figure (but notice that the graph
does indicate the value of energy for $x^{0}_{@B} =0$ which is the case
with Alice as both the emitter and the detector, {\it i.e.} no distance between emitter and detector).
For the case of de Sitter expansion this gives indeed lower values
of detection energy at higher values of detection time.

And Fig.\ref{fig:Lateshift_w_SR} also shows
(solid line) the quantitative behaviour of lateshift produced by momentum-space curvature. Here too
lower values
of detection energy are found at higher values of detection time, but, as here shown in Sec.\ref{sec:kPoincareDetails}
the exponential law governing these correlations takes form dual to the corresponding exponential law
found for the spacetime-curvature case.

\vskip -0.3cm

\section{Two types of relative locality}

A striking aspect of some of the studies triggered by the recent interest in Planck-scale curved
momentum space is the one concerning relative spacetime locality. It is found that in a variety of contexts
the requirement of locality of physical theories must be weakened in presence of momentum-space curvature.
One still insists that events should be local, but allows for the possibility that the locality be manifest
only in the coordinatizations of the event given by nearby observers. The inferences about the event made by distant observers
(according to the coordinatizations they adopt) may not manifest the locality of the event witnessed by nearby observers.
So this weaker principle of locality allows for a relative notion of spacetime locality to replace the ordinary
absolute (observer-indepedent) notion of locality.

In this section we shall summarize our results (later described in greater detail) showing that even just
for free particles on a de Sitter momentum space observers could naturally adopt coordinates such that
relative spacetime locality is present. This will establish a connection between the sort of framework we are
here considering and the presence of relative spacetime locality found in previous studies based
on various aspects of curvature of momentum space, such as \cite{bob,leelaurentGRB,kappabob,transverse,anatomy}.

And still in this section we shall also revisit briefly classic results on the implications
of spacetime curvature which (when analyzed on the background of the recent interest in relative spacetime locality)
deserve to be labelled as effects of ``relative momentum-space locality":
once again there is a duality between the relativity of spacetime locality produced by a de Sitter momentum space
and the relative momentum-space locality produced by a de Sitter spacetime.

\subsection{Relative spacetime locality from momentum-space curvature}\label{antelateshift}

In our simple framework of free particles on a de Sitter momentum space the opportunity
for discussing relative spacetime locality comes from the option of choosing
between the coordinates $x^0,x^1$ and the coordinates $\chi^{0},\chi^{1}$.
 For the discussion of lateshift given in the previous section we relied on the coordinates $x^0,x^1$
 which are free from relative-locality features, but it should be noticed that those coordinates
do not generate translations on momentum space (so they miss one of the defining properties that spacetime
coordinates enjoy when momentum space has no curvature).
The coordinates $\chi^{0},\chi^{1}$ do generate translations on our curved  momentum space
but are affected, as we shall now see, by relative-locality features.

It is useful to
note down again the relationship between $x^0,x^1$ and $\chi^{0},\chi^{1}$  coordinates,
$$\chi^1 = x^1$$
$$\chi^0 = x^0 - \ell x^{1}p_{1}$$
and to also note here  some of the results derived in later sections
when working with the
$x^0,x^1$ coordinates:

\noindent
$\bullet$ with $x^0,x^1$ coordinates the worldlines of massless particles of energy-momentum $p^0,p^1$
have the form
\begin{equation}
x^1-\bar{x}^1 =  \frac{p^1}{|p^1|}e^{\ell p^0} (x^0-\bar{x}^0),\label{primawordlkmink}
\end{equation}

\noindent
$\bullet$  the coordinate transformations between observers Alice and Bob connected by a pure translation
of parameters $a^0,a^1$ are
\begin{equation}
\begin{split}
& x^0_B =  x^0_A-a^0\\
& x^1_B =  x^1_A-a^1.
\end{split}
\label{xtwo}
\end{equation}

 \noindent
 $\bullet$ with $\chi^{0},\chi^{1}$
  coordinates the worldlines of massless particles of energy-momentum $p^0,p^1$
have the form
\begin{eqnarray}
\chi^1-\bar{\chi}^1=\frac{p^1}{|p^1|}(\chi^0-\bar{\chi}^0)\,,
\label{chione}
\end{eqnarray}

\noindent
$\bullet$ and the coordinate transformations between observers Alice and Bob connected by a pure translation
of parameters $a^0,a^1$ are
\begin{equation}
\begin{split}
& \chi^0_B =  \chi^0_A-a^0+a^1\ell p_1\\
& \chi^1_B =  \chi^1_A-a^1.
\label{trasfkmink}
\end{split}
\end{equation}

On the basis of these observations
one already gets a rather
clear picture of the situation: on one side, with
$x^0,x^1$ coordinates,
the curvature of momentum space affects the form of the worldline (producing an energy-dependent velocity)
but leaves the translation transformations unaffected, while on the other side,
with $\chi^{0},\chi^{1}$  coordinates, one has the opposite situation of
worldlines unaffected by the curvature of momentum space but with translation transformations
that reflect momentum space curvature.

As already stressed in previous studies of relative spacetime locality these apparently alternative
pictures of the same physical system are ultimately found to agree on the ``true observables" of such theories
which are times of emission or detection ``at observers" (in the spatial origin of the observer).
But for this consistency for observables obtained within the two alternative coordinatizations
 an important role is played by relative spacetime locality.

To see this it is convenient to contemplate the case of two particles of different energy emitted simultaneously
at Alice toward Bob.
Adopting the $x^0,x^1$ coordinates, with a momentum dependent coordinate velocity, one evidently then finds that
the times of arrival at Bob of the two particles are different. This of course is nothing else but the
lateshift we already described above.
When using the $\chi^{0},\chi^{1}$  coordinates to describe the same situation one finds that according to Alice
the times of arrival at Bob are identical, since the coordinate velocity in the
coordinatization $\chi^{0},\chi^{1}$ is momentum independent.
But with $\chi^{0},\chi^{1}$  coordinates one must take into account relative locality, {\it i.e.}
the fact that translation transformations are affected by momentum space curvature. And the result is that
while according to Alice the particles reach Bob simultaneously, actually according to Bob (an observer near the detections)
the detections are not simultaneous: what Bob finds using $\chi^{0},\chi^{1}$  coordinates
is a difference of detection times  that reproduces exactly the
difference of detection times at Bob obtained with  $x^0,x^1$ coordinates.
So one has the same final result for the lateshift effect using both types of coordinates, even though some
aspects of the analysis do change. The point is that the two choices of coordinatization always agree on
which were the times of emission at Alice according to Alice and which were the detection times at Bob according to Bob.
All this is summarized in Fig.~3.

\begin{figure}[h!]
\centering
\fbox{\includegraphics[scale=0.59]{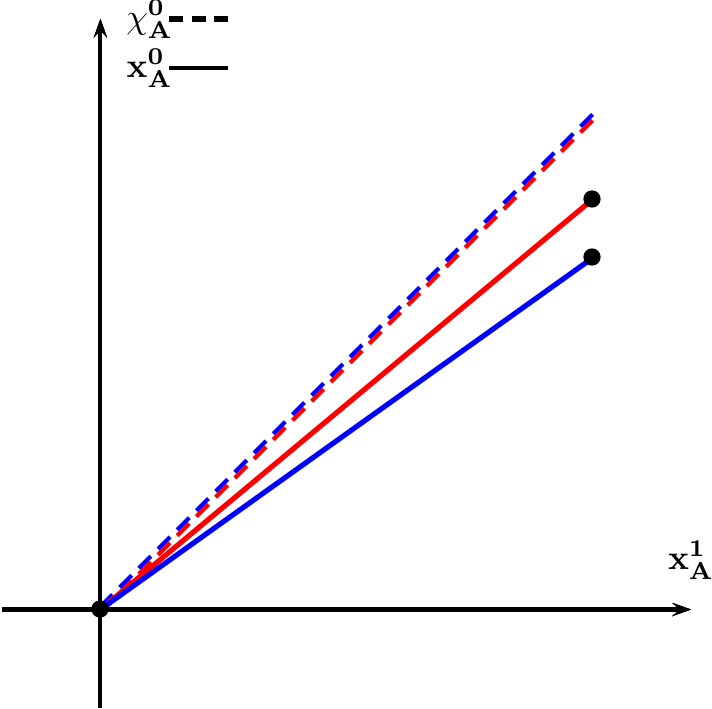}}\\
\fbox{\includegraphics[scale=0.60]{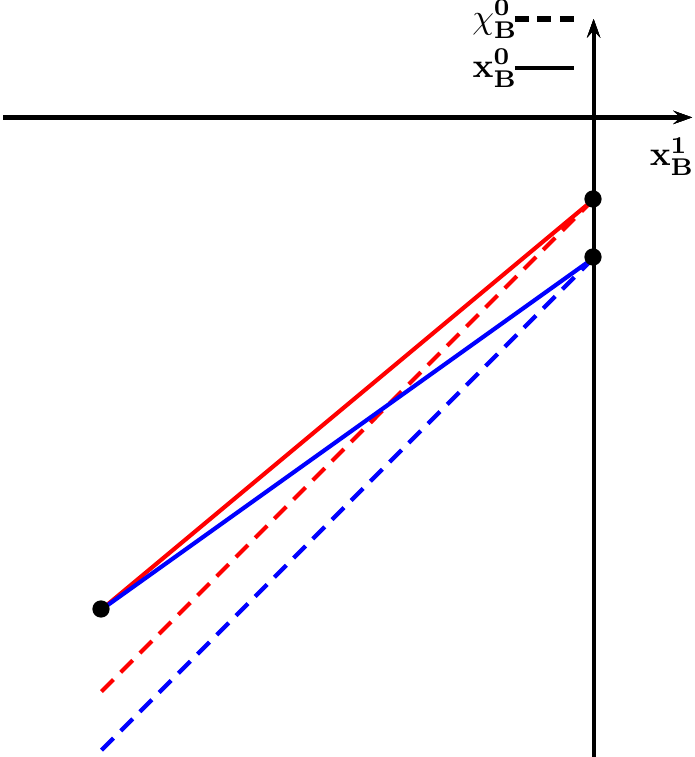}}
\caption{\footnotesize  We here visualize our preferred illustrative example of a relative-spacetime-locality
analysis. Alice's descriptions of massless-particle worldlines are in the top panel, while Bob's descriptions of the
same worldlines are in the bottom panel. Both with $x^0,x^1$ coordinates (solid lines) and with $\chi^{0},\chi^{1}$  coordinates (dashed lines)
Alice has that the
emission is simultaneous and Bob has that the detections occur at different times (same time difference
within both coordinatizations). The points marked on the solid lines identify the values of $\{ p_0 , x^0 \}$ measured at Alice
and at Bob. The peculiarities introduced by
the coordinatization $\chi^{0},\chi^{1}$, affected by relative spacetime locality,
only play a role in the inferences the observers make about distant events: adopting $\chi^{0},\chi^{1}$  coordinates Alice's
would describe the distant detections at Bob as simultaneous and Bob would describe the distant emissions at Alice
as not simultaneous.  Equations in support of these figures will be derived in Sec.\ref{sec:kPoincareDetails}.}
\label{dS-MS-spacetime-B}
\end{figure}

\pagebreak

%The particle's coordinates and momenta in Bob's reference frame %($x_{B}^{\mu},p_{\nu}^{B}$) read in Alice coordinates %($x_{A}^{\mu},p_{\nu}^{A}$):

%\begin{eqnarray}
%&& x^0_{B} = x^0_{A} - a^0 \label{k-transx0}\\
%&& x^1_{B} = x^1_{A} - a^1 \label{k-transx1}\\
%&& p_0^{B} = p_0^{A} \label{k-transp0}\\
%&& p_1^{B} =  p_1^{A} \label{k-transp1}
%\end{eqnarray}

\subsection{Momentum-space relative locality from spacetime curvature}\label{anticipiamoDS-ST}
Our next task is to expose the fact that relative locality is also produced by spacetime curvature:
we find that momentum-space locality is relative when adopting one of the natural choices of
momentum-space coordinates
for the dS-spacetime case.
The choice of coordinatization of momentum space where the curvature of dS spacetime
produces relative-momentum-space-locality effects was here already introduced earlier:
it is the one with coordinatization of momentum space given by $\Pi_0,\Pi_1$,
the conserved charges generating translations of de Sitter spacetime.
We note down again here their relationship to the coordinatization given in terms of energy $p_0$ and spatial
momentum $p_1$, canonically conjugate to the spacetime coordinates:
\begin{equation}
\begin{split}
&\Pi_0=p_0-H x^1 p_1\\
&\Pi_1=p_1
\end{split}
\end{equation}
Again it is useful to characterize the relative-locality effects by considering two observers, Alice and Bob,
on the worldline of a massless particle (the particle crosses the {\underline{spacetime origins}} of both Alice and Bob).
In the dS spacetime two such observers are connected by a translation with translations parameters
linked simply by
$$a^0=a^1$$
For such observers the charges $\Pi_0,\Pi_1$, conserved along the worldline of the massless particle,
must satisfy \footnote{We are specifying our analysis to the case of negative $\Pi_1$ which gives the case
of particle emitted at Alice and propagating toward Bob along the positive $x$-direction.
(The case of positive $\Pi_1$, {\it i.e.} $\Pi_1=\Pi_0$ is  equally interesting, but of course gives rise
to exactly the same qualitative picture.)}
\begin{equation}
\Pi_0=-\Pi_1
\label{pipi}
\end{equation}
But while both Alice and Bob agree on this relation, they attribute different values to these charges:
the translation transformations of dS spacetime act non trivially on the charges $\Pi_0$ and $\Pi_1$,
$$\{\Pi_0,\Pi_1\}=H\Pi_1$$
and the net result is the following relationship between the values attributed to
the charges by Alice and by Bob: \begin{eqnarray}
%\Pi_1^B &=&\sum_{n=0}^{\infty}\frac{1}{n!}\{a^0\Pi_0 +a^1\Pi_1 ,\{..\{a^0\Pi_0 +a^1\Pi_1 ,\Pi_1\}..\}\}=\nonumber\\
%&=& e^{-H a^0} \Pi^A_1\\
%\Pi_0^B &=&\sum_{n=0}^{\infty}\frac{1}{n!}\{a^0\Pi_0+a^1\Pi_1,\{..\{a^0\Pi_0+a^1\Pi_1,\Pi_0\}..\}\}=\nonumber\\
%&=& \Pi^A_0 + H a^1 e^{-H a^0} \Pi^A_1
&& \Pi^B_0 = \Pi^A_0 + \Pi^A_1 \left(1-e^{-H a^0}\right) \label{jocPIa}\\
&& \Pi^B_1 = e^{-H a^0} \Pi^A_1 \label{jocPIb}
\end{eqnarray}
These relationships are the essence of the relative momentum-space locality caused by spacetime
curvature which we here want to highlight.
It is useful to the understanding of this feature to see how the key aspects of the $\Pi_0 , \Pi_1$ coordinatization
get described if instead one uses the more customary  $p_0 , p_1$ coordinatization.
A key point in this respect is that the result $\Pi_0=-\Pi_1$ gets converted into
\begin{equation}
%p_0^2-p_1^2e^{-2H x^0}=0,
p_0=-p_1 e^{-H x^0},
\label{pppp}
\end{equation}
and the transformation  laws (\ref{jocPIa}) and (\ref{jocPIb}) get converted into
\begin{eqnarray}
&& p_0^B =  p_0^A \label{jocMOMa}\\
&& p_1^B =  p_1^A e^{-H a^0} \, . \label{jocMOMb}
\end{eqnarray}
Evidently the key aspect for our analysis of relative momentum-space locality is the
comparison of (\ref{jocPIa}) and (\ref{jocMOMa}) and how those differences
exactly compensate the differences between (\ref{pipi}) and (\ref{pppp}).
To see this it is convenient to contemplate the case of two particles which according to Alice
have the same energy but are emitted at different times
 toward Bob.
Adopting the $p^0,p^1$ coordinates, with trivial transformation (\ref{jocMOMa}) of $p^0$,
 Alice sees redshift as an effect encoded fully in (\ref{pppp}), which in particular is such that
 the energies at Bob are different (because of the different emission times)
 even though the energies at Alice are the same. If Bob also adopts the $p^0,p^1$ coordinates he gets
 a picture completely consistent with Alice's, since with the $p^0,p^1$ coordinates there is no
 relative momentum-space locality.

 If instead Alice and Bob adopt the  $\Pi_0 , \Pi_1$ coordinatization, affected by relative locality,
 the same physical picture is described in a different way.
 Both Alice and Bob have $\Pi_0$ as a conserved charge, but the value Alice gives to $\Pi_0$ is different
 from the value Bob gives to $\Pi_0$, as specified by  (\ref{jocPIa}).
 The mismatch between $\Pi_0$ according to Alice and $\Pi_0$ according to Bob depends on the time-translation
 parameter that connects Alice to Bob, so the fact that the particles are emitted at different times at Alice
 renders them differently subject to redshift.

 The net result is that the pictures given by the $p^0,p^1$ coordinates and by $\Pi_0 , \Pi_1$ coordinates
 are intuitively consistent with each other
  for what concerns the values of energy measured (and the times of those measurements) at both
 Alice and Bob. The only mismatches between the pictures with $p^0,p^1$ coordinates
 and the picture with $\Pi_0 , \Pi_1$ coordinates
 concerns inferences about energies at distant emission/detection events: those inferences are misleading
 when adopting the $\Pi_0 , \Pi_1$ coordinates.
 Alice witnesses emissions of particles with the same $\Pi_0$, but Bob infers that those distant emissions were
 with different $\Pi_0$.
 And similarly  Bob witnesses detections of particles with different $\Pi_0$, but Alice infers that those distant detections
 are at the same $\Pi_0$.\\
All this is summarized in Fig.~4.

\begin{figure}[h!]
\centering
\fbox{\includegraphics[scale=0.52]{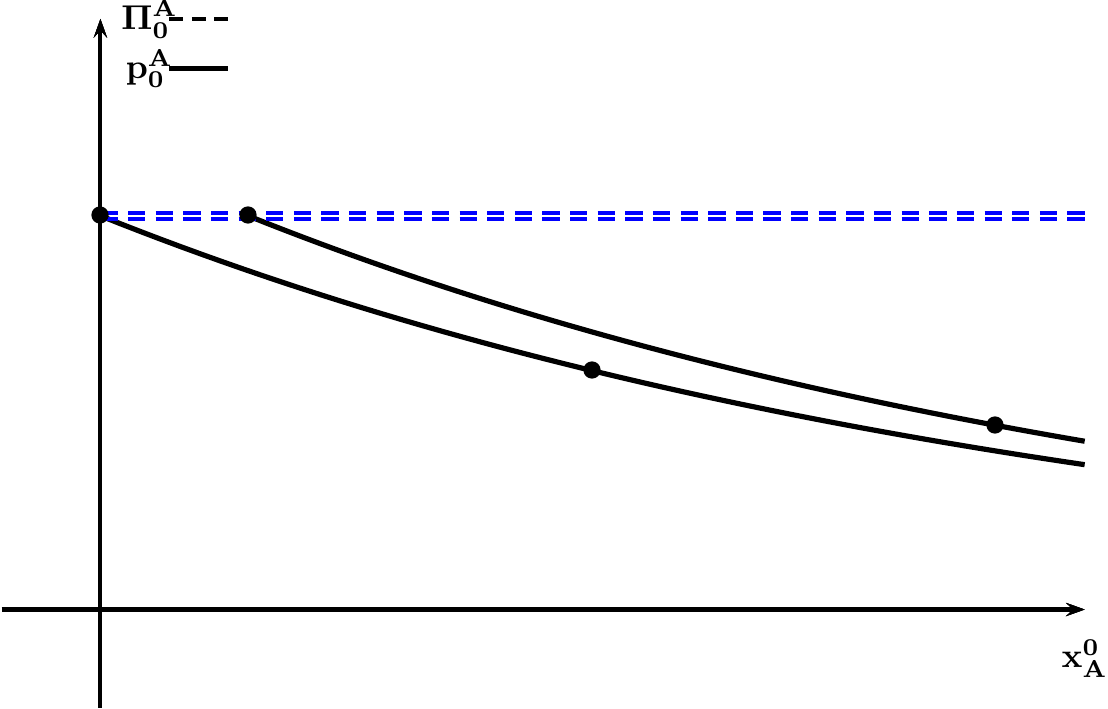}}
\fbox{\includegraphics[scale=0.52]{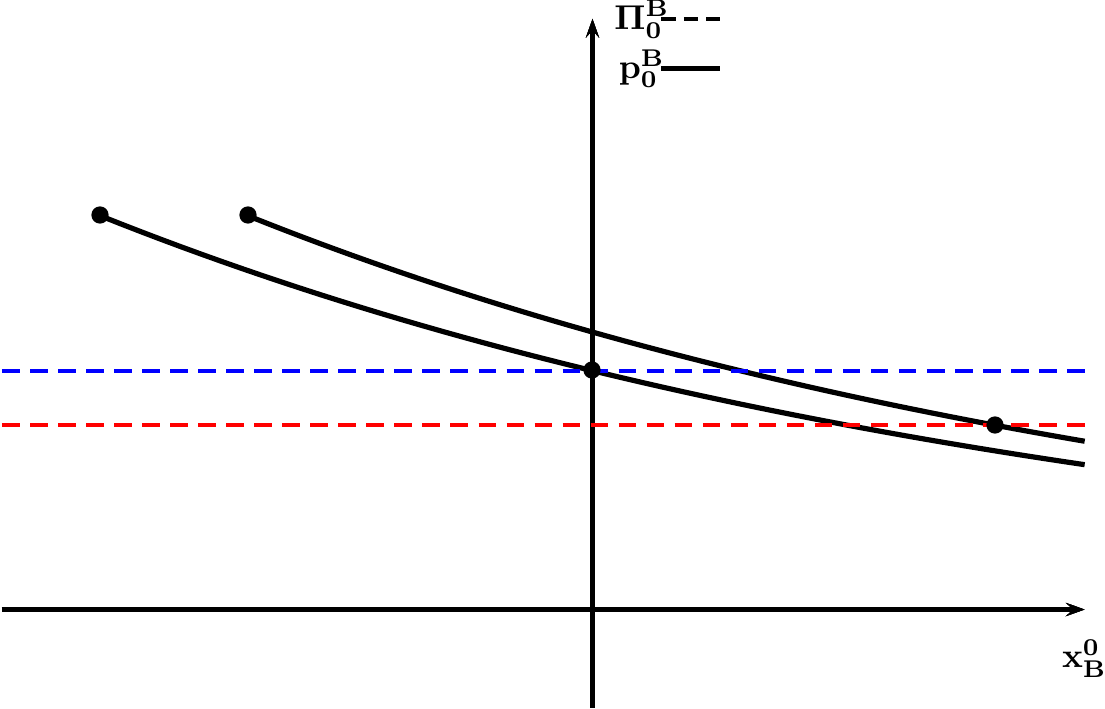}}
\caption{\footnotesize  We here visualize our preferred illustrative example of a relative-momentum-space-locality
analysis, which concerns the evolution in time of $\Pi_{0}$ and $p_0$ on the worldlines of massless particles.
Alice's descriptions of some worldlines are in the top panel, while Bob's description of the same worldlines
are in the bottom panel. Both
with $p_0,p_1$ coordinates (solid lines) and with $\Pi_{0},\Pi_{1}$  coordinates (dashed lines)
Alice has that the
emitted particles have the same energy and Bob has that the energies at detection are different (same energy difference
within both coordinatizations). The peculiarities introduced by
the coordinatization $\Pi_{0},\Pi_{1}$, affected by relative momentum-space locality,
only play a role in the inferences the observers make about distant events: adopting $\Pi_{0},\Pi_{1}$  coordinates Alice
would describe the distant detections at Bob as having the same energy and Bob would describe the distant emissions at Alice
as having different energy. Equations in support of these figures will be derived in Sec.\ref{sec:deSitterDetails}.}
\label{fig:DSBob}
\end{figure}

\newpage

\section{de Sitter space-time}\label{sec:deSitterDetails}

In the previous sections we summarized the key points of our thesis,
establishing a duality between redshift in dS spacetime and lateshift in dS momentum space,
and establishing how this duality also affects the associated relative-locality effects.
As announced we shall now provide more detailed derivations of those key points of our thesis.
We start in this section with the dS-spacetime case.

We start by noticing that in comoving coordinates
the dS-spacetime metric, from which we obtain the spacetime interval (\ref{eq:dSmetric}), and its inverse can be represented by the matrices
\begin{equation}
g_{\mu\nu}=\left(\begin{array}{cc}
1 & 0\\
0 & -e^{2H x^0}
\end{array}\right)\;,\;\;\;g^{\mu\nu}=\left(\begin{array}{cc}
1 & 0\\
0 & -e^{-2H x^0}
\end{array}\right)\label{dsmetrmatr}
\end{equation}
The physical momentum $p^\mu$ of a particle of mass $m$, measured by a free-falling observer, is
$$
p^\mu=m \dot{x}^\mu
$$
where (as we shall do consistently)
the dotted coordinate $\dot{x}^{\mu}$ is differentiated with respect a worldline affine parameter $\tau$,
so that $\dot{x}^{\mu}\equiv dx^{\mu}/d\tau$.
In the following we will mainly work with their lowered-index version,
\begin{equation}
p_{0}=p^{0}, \;\; p_j=-\delta^j_i e^{2H x^0}p^i\;,
\end{equation}
which satisfy ordinary Poisson algebra with coordinates:
\begin{eqnarray}
&\{p_0,x^0\}&=1\;,\;\;\;\{p_0,x^1\}=0\;,\\
&\{p_1,x^0\}&=0\;,\;\;\;\{p_1,x^1\}=1\;,
\end{eqnarray}
We use standard notation for
 Poisson brackets
\begin{equation}
\{A,B\} = \omega_{a b} \frac{\partial A}{\partial \xi^a} \, \frac{\partial B}{\partial \xi^b}
\end{equation}
where $\xi^a$ are the phase space coordinates, and $\omega_{a b}$ identifies the phase space symplectic structure.\\
The conserved charges associated with the translation and boost transformations we used in section \ref{anticipiamoDS-ST} have the following representation on coordinates and momenta:
\begin{eqnarray}
&\Pi_0=p_0-Hx^1p_1\;,\;\; \Pi_1=p_1,\, \label{PiDeSitt}\\ &N=x^1 p_0+\left(\frac{1-e^{-2H x^0}}{2H}-\frac{H}{2}(x^{1})^{2}\right)p_1\label{dsboost1}\,.
\end{eqnarray}
We can also re-express the boost charge in terms of translation-transformation charges $\Pi_\alpha$:
\begin{equation}
N=x^1\Pi_0+\left(\frac{1-e^{-2H x^0}}{2H}+\frac{H}{2}\left(x^{1}\right)^2\right)\Pi_1.\label{dsboost2}
\end{equation}
These charges satisfy the algebra
\begin{eqnarray}
&\{\Pi_0,\Pi_1\}=H\Pi_1\;,\;\;\{N,\Pi_1\}=-\Pi_0&\label{dspi1pi0}\\
&\{N,\Pi_0\}=-\Pi_1+HN\, ,&
\end{eqnarray}
which in particular admits
the following ``mass-Casimir" invariant
\begin{equation}
\mathcal{C}=\Pi_0^2-\Pi_1^2+2HN\Pi_1 \, .\label{dsCasimir}
\end{equation}
The Poisson brackets between conserved translation  charges and coordinates define a symplectic structure
\begin{eqnarray}
&\{\Pi_0,x^0\}&=1\;,\;\;\;\{\Pi_0,x^1\}=-Hx^1\;,\\
&\{\Pi_1,x^0\}&=0\;,\;\;\;\{\Pi_1,x^1\}=1\;,
\end{eqnarray}
which we anticipated in (\ref{phasespaceTWO}).\\
The Casimir relation (\ref{dsCasimir}) for the conserved charges leads to the dS-spacetime mass-shell condition by substituting the expression, (\ref{PiDeSitt}) and (\ref{dsboost1}), of the conserved charges $N$, $\Pi_\alpha$ in terms of the physical momenta $p_\alpha$ and coordinates $x^\beta$:
\begin{equation}
(p_0)^2-(p_1)^2e^{-2H x^0}=m^2 \, . \label{dsreldisp}
\end{equation}
In particular  one has $p_0=|p_1|e^{-Hx^0}$ for massless particles.

Note that we restrict our focus on the case of negative $p_1$ (so that  $\dot{x}^{1}/\dot{x}^0$ is positive) and therefore
 take $p_0= - p_1 e^{-Hx^0}$ as we already showed in (\ref{effectiveonshellness}).\\
The evolution of coordinates $x^\mu$ along a particle worldline with parameter $\tau$ can be obtained in manifestly covariant
form by using a standard Hamiltonian setup with (\ref{dsreldisp}) playing the role of Hamiltonian:
\begin{equation}
\begin{split}
\dot{x}^1&=\hspace{-.1cm}\{(p_0)^2-e^{-2H x^0}(p_1)^2,x^1\}\hspace{-.1cm}=-2e^{-2 H x^{0}}p_1\\
\dot{x}^0&=\hspace{-.1cm}\{(p_0)^2-e^{-2H x^0}(p_1)^2,x^0\}\hspace{-.1cm}=2p_0\hspace{-.1cm}= 2|p_1|e^{-Hx^0}.
\end{split}\label{dsxpunto}
\end{equation}
Then the worldline of a massless particle with initial conditions $x^1(\tau=0)=\bar{x}^1$ and  $x^0(\tau=0)=\bar x^{0}$ reads:
\begin{eqnarray}
x^1(x^0)&-&\bar{x}^1\equiv\int_{\bar{x}^0}^{x^0}\frac{\dot{x}^1}{\dot{x}^0}dx^0=
\left(\frac{e^{-H\bar{x}^0}-e^{-Hx^0}}{H}\right).\;\;\label{dsprocword}
\end{eqnarray}
In particular for $\bar{x}^1=\bar{x}^0=0$ this gives a wordline of the form already anticipated in Eq.(\ref{dS-wl})
of Subsection II.A.

A convenient way to expose the effects of redshift can be based on the comparison of measurements of the same particle
made by two different observers. For this we need to explicitate  the form of finite translation transformations.

The action of the symmetries generators on the phase space functions is represented through the ordinary left action of Lie groups
\begin{equation}
G_\mathbf{v} \rhd A(\xi) = \sum^\infty_{n=0} \frac{1}{n!} \underbrace{\{\mathbf{v}\cdot \mathfrak{g}, \dots \{\mathbf{v}\cdot \mathfrak{g}}_{n \, \text{times}}, A(\xi) \} \dots \}.
\end{equation}
being $G_\mathbf{v}$ an element of the group identified by the vector parameter $\mathbf{v}$, and $\mathfrak{g}$ a set of  elements of the algebra (the generators of the symmetry transformation associated to the group element $G_v$). For example a generic translation $\mathcal{T}_a$, connected to the time and space translation generators $\vec{\mathfrak{t}}=(\mathfrak{t}_0,\mathfrak{t}_1)$ by the translation parameters $\vec{a}=(a^0,a^1)$, acts on a generic phase space function $F(\xi)$ as
\begin{equation}
\mathcal{T}_a \rhd F(\xi) = \sum^\infty_{n=0} \frac{1}{n!} \underbrace{\{\vec{a}\cdot \vec{\mathfrak{t}}, \dots \{\vec{a}\cdot \vec{\mathfrak{t}}}_{n \, \text{times}}, F(\xi) \} \dots \}.
\end{equation}
Then, translation transformations $ \mathcal{T}_a$ in deSitter spacetime act in the following way on spacetime coordinates and physical momenta:
\begin{equation}
\begin{split}
%p_0^B&=&\sum_{n=0}^{\infty}\frac{1}{n!}\{-a^0\Pi_0-a^1\Pi_1,\{...\{-a^0\Pi_0-a^1\Pi_1,p_0\}\}\}=\nonumber\\
%&=& p_0^A\\
%p_1^B&=&\sum_{n=0}^{\infty}\frac{(-a^0)^n}{n!}\{\Pi_0,\{...\{\Pi_0,p_1\}\}\}=p_1^A e^{-a^0H}\\
%x_B^1&=&\sum_{n=0}^{\infty}\frac{1}{n!}\{-a^0\Pi_0-a^1\Pi_1,\{...\{-a^0\Pi_0-a^1\Pi_1,x^0\}\}\}=\nonumber\\
%&=&x_A^1 e^{a^0H}-a^1 e^{a^0H}\\
%x^0_B&=&\sum_{n=0}^{\infty}\frac{(-a^0)^n}{n!}\{\Pi_0,\{...\{\Pi_0,x^0\}\}\}=x^0_A-a^0
 &p_0^B = \mathcal{T}_a \rhd p_0^A = p_0^A\\
 &p_1^B = \mathcal{T}_a \rhd p_1^A = p_1^A e^{-a^0H} \\
& x^0_B = \mathcal{T}_a \rhd x^0_A = x^0_A-a^0\\
& x^1_B = \mathcal{T}_a \rhd x^1_A = e^{a^0 H} \left(x^1_A - \frac{a^1}{a^0}\frac{1-e^{-a^0 H}}{H}\right)
\end{split}\label{eq:deSitterTranslations}
\end{equation}
where $a^{0}$ and $a^{1}$ are, respectively, time and space translation parameters connecting the two observers.
Therefore, if an observer Alice observes the following worldline of a photon emitted at her origin $(\bar{x}^0_A=0,\bar{x}^1_A=0)$:
\begin{equation}
x^1_A(x^0)=\left(\frac{1-e^{-Hx^0}}{H}\right),
\end{equation}
then a second observer Bob, connected to Alice by a translation transformation, will observe the worldline
\begin{eqnarray}
%&&x^1_B(x^0)-\bar{x}^1_B=\frac{p_B^1}{p^0_B}\left(\frac{e^{-H\bar{x}^0_{B}}-e^{-Hx^0}}{H}\right)\Rightarrow\nonumber\\
%&&\Rightarrow
 x^1_B(x^0)+\frac{a^1}{a^0}\frac{ e^{a^0H}-1}{H}=\left(\frac{e^{Ha^0}-e^{-Hx^0_{B}}}{H}\right).
\end{eqnarray}
From this one sees that the family of observers reached in their spacetime origin by the signal emitted in Alice's
spacetime origin  $(x^{1}_{B}(x^{0}_{B}=0)=0)$ are the ones connected to Alice by
translations whose translation parameters obey the following simple relation
\begin{equation}
a^1=a^0.
\label{dS-ST-Bob*}
\end{equation}
We are at this point equipped to also highlight the aspects of relative momentum-space locality, {\it i.e.}
the misleading inferences that can arise when the dS-spacetime observers adopt the $\Pi_0 , \Pi_1$ coordinatization
of momentum space.
A key aspect of this is due to the fact that an emitter/observer Alice who measures a certain value of $p_0$
(determines $p_0$ ``at Alice")
also determines a corresponding value of $\Pi^A_0$ with $\Pi^A_0 = p_0$, and $\Pi_0$ is a conserved charge so
the value Alice assigns to $\Pi^A_0$ at some distant detector Bob still is given by the value
of  $p_0$ ``at Alice".
But translations, while  acting trivially on $p_0$ as we saw in Eqs. (\ref{jocMOMa}) and (\ref{eq:deSitterTranslations}) (but $p_0$ is not conserved along the worldline), act non-trivially on $\Pi^A_0$, so Alice's inference for the value of $\Pi_0$ at Bob actually disagrees
from what Bob determines for $\Pi_0$:
\begin{eqnarray}
 \Pi^B_0 &=& \mathcal{T}_a \rhd \Pi^A_0 = \Pi^A_0 +  \frac{a^1}{a^0}\Pi^A_1\left( 1-e^{-H a^0}\right)\label{relazioniPiBA1} \\
 \Pi^B_1 &=& \mathcal{T}_a \rhd \Pi^A_1 =e^{-H a^0} \Pi^A_1 \, , \label{relazioniPiBA2}
\end{eqnarray}
as we already anticipated in (\ref{jocPIa}) and (\ref{jocPIb}). From these equations it follows that
\begin{equation}
\Pi_0^B=e^{-a^0 H}\Pi_0^A \, .\label{redshiftPi}
\end{equation}
Notice that all this is fully consistent with the fact that for both observers the value of $\Pi_0$ (conserved on the worldline)
coincides with the value of $p_0$ when the worldline crosses the observer's spacetime origin: $\Pi_0 = p_0\Big|_{x^0=0,x^1=0}$. Moreover of course both observers agree (though giving different values to these charges)
with the relationship $\Pi_0=-\Pi_1$ we showed in (\ref{pipi}) among charges on worldlines that cross their spacetime origin.

This reproduces the conceptual picture of dS spacetime given in the previous sections.
In order to also confirm the quantification of effects given in the previous sections
we can now consider the case in which two massless particles are emitted at Alice,
one at time $\bar x^{0}_{A}=0$ and the other one at time $\tilde{x}^{0}_{A}\,=\Delta x^{0}_{A}$,
both with energy $p_{0}^{A}$.
As second observer, Bob, we take one such that the first particle worldline crosses the osberver's spacetime origin,
 so that this particle's worldline is described in Bob's coordinatization by
\begin{equation}
%x^{1}_{B}-\bar x^{1}_{B}=\frac{e^{-H \bar x^{0}_{B}}-e^{-H x^{0}_{B}}}{H} \Rightarrow
x^{1}_{B}=\frac{1-e^{-H x^{0}_{B}}}{H},
\end{equation}
where we have used the relations (\ref{eq:deSitterTranslations}) and (\ref{dS-ST-Bob*}).
The  second particle crosses Bob's spatial origin at a time $\tilde{x}^{0}_{B}\,\neq 0$ and Bob describes its worldline as
follows:
\begin{equation}
\tilde{x}^{1}_{B}\,+\frac{a^1}{a^0}\frac{ e^{a^0H}-1}{H}=\frac{e^{-H(\Delta x^{0}_{A}-a^{0})}-e^{-H \tilde{x}^{0}_{B}\,}}{H}.
\end{equation}
Using again eq. (\ref{dS-ST-Bob*}) this can be written in the form:
\begin{equation}
\tilde{x}^{1}_{B}\,=\frac{e^{-H \Delta x^{0}_{B}}-e^{-H \tilde{x}^{0}_{B}\,}}{H},
\end{equation}
with
$$\Delta x^{0}_{B} \equiv -\frac{\ln\left(-1+e^{Ha^{0}}+e^{H(a^{0}-\Delta x^{0}_{A})}\right)}{H} \, .$$

Especially for what concerns the values determined for  energies it is useful to offer our analysis
using a very explicit notation, capable of differentiating among different ways in which a certain observable
could be determined: for quantities measured at Bob's spatial origin adopting  Bob's coordinatization
we use the subscripts $B@B$; for quantities described within Alice's  coordinatization but concerning inferences
for values of observables at Bob's spatial origin we use the subscripts $A@B$;
for quantities measured at Alice's spatial origin adopting  Alice's coordinatization
we use the subscripts $A@A$; for quantities described within Bob's  coordinatization but concerning inferences
for values of observables at Alice's spatial origin we use the subscripts $B@A$.

Equipped with this notation
we can quickly assess the situation.
For the first particle, crossing Bob's spatial origin at time $x^{0}_{B@B}=0$, we have the relation we explicitely showed in Fig. \ref{fig:Redshift_w_SR}:
\begin{equation}
p_{0}^{B@B}=p_{0}^{A@B}=e^{-H x^{0}_{A@B}}p_{0}^{A@A} \, ,\label{redshiftinfig}
\end{equation}
where the first equality holds because of the triviality of the action of translation transformations
 on $p_{0}$ and the second equality follows from (\ref{dsreldisp}).
For the second particle, crossing Bob's spatial origin at time $\tilde{x}^{0}_{B@B}\,=\Delta x^{0}_{B}$, we have

\begin{equation}
 \tilde{p}_{0}^{B@B}=\tilde{p}_{0}^{A@B}=e^{-H \tilde{x}^{0}_{A@B}\,}\tilde{p}_{0}^{A@A}=e^{-H \tilde{x}^{0}_{A@B}\,}{p}_{0}^{A@A},\label{eq:redshift}
\end{equation}
where the last equality reflects our choice to focus on two particles with the same energy
at Alice $\tilde{p}_{0}^{A@A}\,=p_{0}^{A@A}$.

Comparing the expression for $p_{0}^{B@B}$ and $\tilde{p}_{0}^{B@B}\,$, as anticipated in (\ref{redshiftmaster}), we find:
\begin{equation}
\tilde{p}_{0}^{B@B}\,=p_{0}^{B@B}e^{-H(\tilde{x}^{0}_{A@B}\,-x^{0}_{A@B})}
\end{equation}
which evidently reproduces the last equation of Subsec.~\ref{sec:redshift}, since
dS-spacetime translations are such that ${p}_{0}^{B@B} = {p}_{0}^{A@B}$,
$\tilde{p}_{0}^{B@B} = \tilde{p}_{0}^{A@B}$
and $\tilde{x}^{0}_{A@B}\,-x^{0}_{A@B}=\tilde{x}^{0}_{B@B}\,-x^{0}_{B@B}=\Delta x^{0}_{B}$.

\section{de Sitter momentum space}
\label{sec:kPoincareDetails}

Our next task is to analyze the dual picture of dS momentum space.
We start by describing the metric on dS momentum space, with the same structure of
the metric on dS spacetime considered in the preceding section.
So we have, as already noted in (\ref{metricDSMS}),
\begin{equation}
dk^2=(dp_0)^2-e^{2\ell p_0}(dp_1)^2 \, ,
\end{equation}
and  in matrix form
\begin{equation}
\zeta^{\alpha\beta}=\left(\begin{array}{cc}
1&0\\
0&-e^{2\ell p_0}
\end{array}\right)\;,\;\;\;\zeta_{\alpha\beta}=\left(\begin{array}{cc}
1&0\\
0&-e^{-2\ell p_0}
\end{array}\right).
\end{equation}

For the coordinatization of spacetime
in this case allowing for curvature of momentum space we find convenient to start with
the possibility of spacetime coordinates $\chi^\mu$
which generate translations on dS momentum space (in analogy with the $\Pi_\mu$ coordinatization of
momentum space adopted for parts of our analysis of properties of dS spacetime). For these we have that
\begin{equation}
\{\chi^0,\chi^1\}=\ell\chi^1.\label{kMink}
\end{equation}
And we shall describe spacetime symmetries of this case with dS momentum space in terms
of charges/generators of space translation, time translation and boost governed by the following
Poisson brackets\footnote{Note that rules of action of boosts on momenta of the type here given in Eq.(\ref{kPoinc})
have been independently of interest in the literature on the $\kappa$-poincar\'e
Hopf algebra\cite{majruegg,lukruegg}, which indeed in one of the formalisms forwhich a connection
with the possibility of dS momentum space had been made\cite{kowadesitter,gacdsrrev2010}.}:
\begin{eqnarray}
 \{p_{1},p_{0}\}&=&0, \label{eq:momentaPoisson} \\
  \{\mathcal{N},p_0\}&=&p_1, \;\;
 \{\mathcal{N},p_1\}=\frac{1-e^{-2\ell p_0}}{2\ell}-\frac{\ell}{2}(p_1)^2.\label{kPoinc}
\end{eqnarray}
These phase-space Poisson brackets are compatible with the Jacobi identities upon assuming that the Poisson brackets
involving $\chi^\mu$ and $p_\mu$ satisfy (also see (\ref{dssimpl}))
\begin{equation}
\begin{split}
&\{p_1,\chi^1\}=-1\;,\;\;\;\{p_1,\chi^0\}=\ell p_1, \noindent\\
&\{p_0,\chi^1\}=0\;,\;\;\;\{p_0,\chi^0\}=-1. \label{kpoinsimpl}
\end{split}
\end{equation}

Just like in the dS-spacetime case the possibilities $\Pi_\mu$ and $p_\mu$  for coordintizing momentum
space are comparably (though complementarily) convenient, for the dS-momentum-space case which we are now considering
one can conveniently coordinatize spacetime either with the coordinates $\chi^\mu$ or with the following
coordinates $x^\mu$:
\begin{eqnarray}
&&x^1\equiv \chi^1\;,\;\;\; x^0\equiv \chi^0+\ell\chi^1 p_1,\label{Smolcoord}\\
&&\{x^{\mu},x^{\nu}\}=0\\
&&\{p_{\mu},x^{\nu}\}=-\delta_{\mu}^{\nu} \label{trasljoc}
\end{eqnarray}
The convenience of these spacetime coordinates resides mainly in the fact that
translation transformations act trivially on them, as shown in (\ref{trasljoc}).

Note that all the phase-space  relations (\ref{kMink}), (\ref{eq:momentaPoisson}), (\ref{kpoinsimpl}) and (\ref{trasljoc})
here given
for the case of dS momentum space
 are dual to the ones, shown in the previous section, that hold in the dS-spacetime case for
 conserved charges $\Pi_{\alpha}$ and spacetime coordinates $x^{\alpha}$: they are  obtained one from the other  through the substitutions  $H\leftrightarrow \ell$, $x^\mu \leftrightarrow p_\mu$ and $\Pi_{\mu}\leftrightarrow \chi^{\mu}$.

In comparing results obtained with the two coordinatizations of spacetime suitable for
theories with dS momentum space, $\chi^\mu$ and $x^\mu$, it can be useful to also take notice
of the following two possible representations of our boost generator:
\begin{equation}
{\cal N}=p_1\chi^0+\left(\frac{1-e^{-2\ell p_0}}{2\ell}+\frac{\ell}{2}p_1^2\right)\chi^1
\end{equation}
\begin{equation}
{\cal N}=p_1 x^0+\left(\frac{1-e^{-2\ell p_0}}{2\ell}-\frac{\ell}{2}p_1^2\right) x^1.
\end{equation}

Once again the duality of these formulae with the representations (\ref{dsboost1}) and (\ref{dsboost2}) of the boosts on dS spacetime
 is easily seen through the exchange $\Pi_\alpha\leftrightarrow\chi^\alpha$, $p_\beta\leftrightarrow x^\beta$ and $H\leftrightarrow\ell$.

The mass-Casimir invariant of the algebra (\ref{kPoinc}) is
\begin{equation}
{\cal C}_\ell=\left(\frac{2}{\ell}\sinh\left(\frac{\ell p_0}{2}\right)\right)^2-e^{\ell p_0}p_1^2 \, . \label{kPCasimir}
\end{equation}
Therefore for a massless particle on the dS momentum space one has the on-shell (${\cal C_\ell}=0$) condition
 of the form (\ref{onshellness}):
\begin{equation}
p_1(p_0)=\frac{1-e^{-\ell p_0}}{\ell}\label{kPoinwordmom}
\end{equation}

There have been several studies (see, {\it e.g.}, Refs.\cite{leelaurentGRB,bob,kappabob})
of the implications of momentum-space curvature for the properties of spacetime worldlines
and travel times. While these previous studies focused on results applicable at leading order in $\ell$,
drawing from the strength of the duality here exposed we are now in position
to do an analogous study of the implications of dS momentum space to all orders in  $\ell$ (exact).

The dependence of a  coordinates $\chi^\alpha$ on the  worldline parameter $\tau$ can be again
found using the condition of on-shellness(\ref{kPCasimir}) as Hamiltonian: $\frac{d\chi^\mu}{d\tau} \equiv \dot{\chi}^\mu=\{\mathcal{C}_\ell,\chi^\mu\}$. This leads to
\begin{eqnarray}
 \dot{\chi}^0% &=&\{\Big( \frac{2}{\ell} \sinh(\frac{\ell p_0}{2}) \Big)^2 - e^{\ell p_0} (p_1)^2,x^0\}= \nonumber\\
&=&\frac{1}{\ell}\left(e^{\ell p_0}-e^{-\ell p_0}\right)+\ell p_1^2 e^{\ell p_0}\nonumber\\
 \dot{\chi}^1%&=&\{\Big( \frac{2}{\ell} \sinh(\frac{\ell p_0}{2}) \Big)^2 - e^{\ell p_0} (p_1)^2,\chi^1\}=\nonumber\\
&=&2 \, p_1 \, e^{\ell p0} \, ,\nonumber
\end{eqnarray}
which evidently implies
\begin{eqnarray}
\chi^0\left(\tau\right)&=&\bar{\chi}^0+\left(\frac{1}{\ell}\left(e^{\ell p_0}-e^{-\ell p_0}\right)+\ell p_1^2 e^{\ell p_0}\right)\tau\ ,\nonumber \\
\chi^1\left(\tau\right)&=&\bar{\chi}^1+\left(2 p_1 e^{\ell p0}\right)\tau\ ,\nonumber \end{eqnarray}
where $\bar{\chi}^{\mu}$ are the initial conditions.

Specializing to the case of massless particles and eliminating the parameter $\tau$
we find a wordline of the form (\ref{chione}):
\begin{equation}
\chi^1=\bar{\chi}^1-(\chi^0-\bar{\chi}^0)\ .\label{wordchi}\end{equation}
which is independent on the particle's energy and  momentum.
Notice however that
translations on coordinates act non trivially on both the coordinates $\chi^\mu$  and the momenta, as shown already in (\ref{trasfkmink}):
\begin{eqnarray}
%p_0^B&=&\sum_{n=0}^{\infty}\frac{1}{n!}\{a^0p_0+a^1p_1,\{..\{a^0p_0+a^1p_1,p_0\}..\}\}=\nonumber\\
%&=& p_0^A\\
%p_1^B&=&\sum_{n=0}^{\infty}\frac{1}{n!}\{a^0p_0+a^1p_1,\{..\{a^0p_0+a^1p_1,p_1\}..\}\}=\nonumber\\
%&=& p_1^A\\
%\chi^1_B&=&\sum_{n=0}^{\infty}\frac{1}{n!}\{a^0p_0+a^1p_1,\{..\{a^0p_0+a^1p_1,\chi^1\}..\}\}=\nonumber\\
%&=& \chi^1_A-a^1\\
%\chi^0_B&=&\sum_{n=0}^{\infty}\frac{1}{n!}\{a^0p_0+a^1p_1,\{..\{a^0p_0+a^1p_1,\chi^0\}..\}\}=\nonumber\\
%&=& \chi^0_A-a^0+a^1\ell p_1
&& p_0^B = \mathcal{T}_a \rhd p_0^A = p_0^A\\
&& p_1^B = \mathcal{T}_a \rhd p_1^A = p_1^A\\
&& \chi^0_B = \mathcal{T}_a \rhd \chi^0_A = \chi^0_A-a^0+a^1\ell p_1\\
&& \chi^1_B = \mathcal{T}_a \rhd \chi^1_A =  \chi^1_A-a^1
\end{eqnarray}
where again we denote by $a^{0}$ and $a^{1}$ the translation parameters connecting two different observers.
From this we see that
if Alice observes worldlines of the form (\ref{wordchi})
\begin{eqnarray}
\chi^1_A -\bar{\chi}^1_A=-(\chi^0_A-\bar{\chi}^0_A)\ .
\end{eqnarray}
 for a particle emitted at a point $(\bar{\chi}_A^0=\bar{\chi}^0,\bar{\chi}_A^1=\bar{\chi}^1)$, then a distant observer
 Bob will  observe:
\begin{eqnarray}
\chi^1_B\left(p,\chi^0\right)-\bar{\chi}^1_B=-(\chi^0_B-\bar{\chi}^0_B)\,,
\end{eqnarray}
with $\bar{\chi}^1_B=\bar{\chi}^1-a^1$ and $\bar{\chi}^0_B=\bar{\chi}^0-a^0+a^1\ell p_1$.\\
So when using $\chi^\mu$ coordinates one has that
the form of the worldline is energy-independent but the translation transformation is momentum dependent.\\

It is interesting to compare these findings to the ones using the coordinatization $x^\mu$.
The description of the worldlines in terms of the $x^\mu$ is found by observing that
\begin{eqnarray}
 \dot{x}^0 &=& \{\mathcal{C}_\ell,x^0\}%=\{\Big( \frac{2}{\ell} \sinh(\frac{\ell p_0}{2}) \Big)^2 - e^{\ell p_0} (p_1)^2,x^0\}= \nonumber\\
=\frac{1}{\ell}\left(e^{\ell p_0}-e^{-\ell p_0}\right)-\ell p_1^2 e^{\ell p_0}\nonumber\\
 \dot{x}^1 &=& \{\mathcal{C}_\ell,x^1\}%=\{\Big( \frac{2}{\ell} \sinh(\frac{\ell p_0}{2}) \Big)^2 - e^{\ell p_0} (p_1)^2,x^1\}=\nonumber\\
= 2 \, p_1 \, e^{\ell p0}.\nonumber
\end{eqnarray}
From this, by
integrating $\dot{x}^1$ on the worldline parameter $\tau$, we find
\begin{eqnarray}
 \int^{\tau}_{\tau_0} \dot{x}^1 d\tau   = \int^{x^0}_{\bar{x}^0} \frac{\dot{x}^1}{\dot{x}^0} dx^0 = -e^{\ell p0} (x^0-\bar{x}^0)
\end{eqnarray}
from which is follows that
\begin{equation}
x^1-\bar{x}^1 =  -e^{\ell p0} (x^0-\bar{x}^0),\label{wordliics}
\end{equation}
This shows that with the $x^\mu$ coordinates the form of the worldline of a massless particle
 is momentum dependent, and confirms  Eq.(\ref{primawordlkmink}), which in the case of a particle emitted in the observer's
origin, $\bar{x}^0=\bar{x}^1=0$, reduces to Eq.(\ref{kM-wl}).
As we already noted in Eq.(\ref{xtwo}), in the $x^\mu$ coordinates translations act trivially:
 \begin{eqnarray}
%p_0^B&=&\sum_{n=0}^{\infty}\frac{1}{n!}\{a^0P_0+a^1P_1,\{..\{a^0P_0+a^1P_1,p_0\}..\}\}=\nonumber\\
%&=& p_0^A\\
%p_1^B&=&\sum_{n=0}^{\infty}\frac{1}{n!}\{a^0P_0+a^1P_1,\{..\{a^0P_0+a^1P_1,p_1\}..\}\}=\nonumber\\
%&=& p_1^A\\
%x^1_B&=&\sum_{n=0}^{\infty}\frac{1}{n!}\{a^0P_0+a^1P_1,\{..\{a^0P_0+a^1P_1,x^1\}..\}\}=\nonumber\\
%&=& x^1_A-a^1\\
%x^0_B&=&\sum_{n=0}^{\infty}\frac{1}{n!}\{a^0P_0+a^1P_1,\{..\{a^0P_0+a^1P_1,x^0\}..\}\}=\nonumber\\
%&=& x^0_A-a^0.
&& p_0^B = \mathcal{T}_a \rhd p_0^A = p_0^A\\
&& p_1^B = \mathcal{T}_a \rhd p_1^A = p_1^A\\
&& x^0_B = \mathcal{T}_a \rhd x^0_A =  x^1_A-a^1\\
&& x^1_B = \mathcal{T}_a \rhd x^1_A = x^0_A-a^0
\end{eqnarray}
Then is easy to obtain that if Alice  observes a worldline of the form
\begin{equation}
x^1_A-\bar{x}^1_A =  -e^{\ell p_0^A} (x^0_A-\bar{x}^0_A),
\end{equation}
the translated observer Bob will  agree about the wordline expression in his coordinates,
\begin{equation}
x^1_B-\bar{x}^1_B =  -e^{\ell p_0^B} (x^0_B-\bar{x}^0_B).
\end{equation}
with $\bar{x}^1_B=\bar x^{1}_{A}-a^{1}$ and $\bar{x}^0_B=\bar x^{0}_{A}-a^{0}$.

Summarizing the issue of the choice of spacetime coordinates we have that the form of the worldline
of a massless particle is momentum independent when using the $\chi^\mu$ coordinates
whereas it is momentum dependent when using the
$x^\mu$ coordinates. But this difference is balanced by the other difference:
translations transformations are momentum dependent when using the $\chi^\mu$ coordinates
whereas they are momentum independent when using the
$x^\mu$ coordinates.

Following again the logical line of the previous section, we also observe
that the observers whose origin is crossed by a given massless particle's worldline must be connected
(if in relative rest) by a translation with parameters $a^1$,$a^0$ linked by
\begin{equation}
a^1=-e^{\ell p_0}a^0. \label{eq:kBobStar}
\end{equation}
As in the previous section for dS spacetime,
also in this dS-momentum-space case we are interested in comparing observations made by two observers connected
by a translation transformations. While in the dS-spacetime case it proved useful to consider two particles
emitted at different times with same energy, we find useful for the dS-momentum-space case
to consider two particles emitted simultaneously with different energies.

So let us consider two massless particles emitted with  different energies $p_0$ and $\tilde{p}_0$, in the origin of the observer Alice:
\begin{eqnarray}
&& x^1_{A} = -e^{\ell p_0} x^0_{A} \label{wl-p0-Aqui}\\
&& x^1_{A} = -e^{\ell \tilde{p}_0} x^0_{A} \label{wl-p0prime-Aqui}
\end{eqnarray}
Note that in the dS-momentum-space case the momenta are conserved along the motion
and under invariant under translations, so we omit observer's indices for them in this section.\\
For a translated observer Bob such that the  worldline energy $p_0$  intercepts his origin one has a description
 of the worldlines in terms of the following equations:
\begin{equation}
 x^1_{B} = -e^{\ell p_0} x^0_{B}
\label{wl-p0-Bqui}
\end{equation}
and
\begin{equation}
 \tilde{x}^1_{B}\, = -e^{\ell \tilde{p}_0} \tilde{x}^0_{B}\,+e^{\ell p_{0}}a^{0}(1-e^{\ell (\tilde{p}_{0}-p_{0})})
\label{wl-p0prime-Bqui}
\end{equation}
where we made use of the relation (\ref{eq:kBobStar}).
So the particle with energy $\tilde{p}_{0}$ arrives at Bob's spatial origin at time:
\begin{equation}
\tilde{x}^{0}_{B}\,( \tilde{x}^1_{B}\, =0)=- a^{0}(1-e^{-\ell (\tilde{p}_{0}-p_{0})})
\end{equation}

As done at the end of the previous section, let us use again notation specifying the value of observables
either at Alice or at Bob, and according to Alice's coordinatization of Bob's coordinatization.
Accoding to Alice's coordinatization the particle of energy $p_0$ arrives in Bob's origin at time
\begin{equation}
x^{0}_{A@B}=a^{0}
\label{joczzzpre}
\end{equation}
whereas the other particle is at Bob at time
\begin{equation}
\tilde{x}^{0}_{A@B}\,=a^{0}e^{-\ell (\tilde{p}_{0}-p_{0})}.
\label{joczzz}
\end{equation}
In the particularly interesting case in which the energy $p_{0}$ is small enough that (within a given experimental sensitivity)
the term $e^{\ell p_{0}}$ in (\ref{joczzz})
can be neglected (so the behaviour of that massless particle is as if momentum space was flat
and ordinary special relativity was applicable), then one finds that (\ref{joczzz})
takes the form used for Fig. \ref{fig:Lateshift_w_SR}:
\begin{equation}
\tilde{x}^{0}_{@B}\,=a^{0}e^{-\ell \tilde{p}^{@B}_{0}},\label{lateshiftinfig}
\end{equation}
For completeness let us also observe that from  (\ref{joczzzpre}) and (\ref{joczzz})
it follows that in Alice's coordinatization the relation between the two arrival times at Bob's spatial origin is
\begin{equation}
\tilde{x}^{0}_{A@B}\,=x^{0}_{A@B}e^{-\ell (\tilde{p}_{0}-p_{0})},
\end{equation}
which is the equation (\ref{lateshiftmaster}) reported at the end of section \ref{sec:lateshift}. And we also
stress again that this is completely analogous to Eq. (\ref{eq:redshift}) for the dS-spacetime case,
describing the relation between the energies
inferred by Alice as values of energies at Bob for the case of two massless particles emitted  with same energy
but at different times.

\hspace{.5cm}

\section{Aside on the Newton-Wigner observable}
While, as shown above, it is not necessary for exposing the duality which was here
of interest, there are interesting implicit roles in our analysis for the  Newton-Wigner observable $\mathcal{A}$,
defined by
\begin{equation}
\mathcal{A}=\hspace{-.1cm}\int dx^1-\dot{x}^1d\tau=\int dx^1-\frac{\dot{x}^1}{\dot{x}^0}dx^0.
\end{equation}
Let us start noticing that in the dS-spacetime case for massless particles
 the Hamiltonian constraint can be written as
\begin{equation}
{\cal H}_H=p_1+p_0 e^{H x^0}\label{CH} \, ,
\end{equation}
which in turn allows one to write the  Newton-Wigner observable straightforwardly using (\ref{dsxpunto}) and (\ref{effectiveonshellness}):
\begin{equation}
{\cal A}_H=\int_{0}^{x^0} dx^1-e^{-H x^0}dx^0=x^1-\left(\frac{1-e^{-H x^0}}{H}\right)\label{AH}.
\end{equation}

Of course the Newton-Wigner observable commutes with the Hamiltonian constraint: $\{{\cal H}_H,{\cal A}_H\}=0$. \\

In the dS-momentum-space case we have that  the Hamiltonian constraint
takes the form  (\ref{kPCasimir}),
\begin{equation}
{\cal H}_\ell=p_1-\frac{1-e^{-\ell p_0}}{\ell}\label{Cell},
\end{equation}
from which it follows that the  Newton-Wigner observable
can be written as
\begin{equation}
{\cal A}_\ell=\int_{0}^{x^0} dx^1-e^{\ell p^0}dx^0=x^1+e^{\ell p^0}x^0\label{Aell}.
\end{equation}
(And again on easily verifies that the Newton-Wigner observable commutes with Hamiltonian constraint:$\{{\cal H}_\ell,{\cal A}_\ell\}=0$.)\\

A very efficient way to summarize the duality we here exposed between the dS-spacetime case and the dS-momentum-space case
is contained in Eqs. (\ref{Cell}), (\ref{Aell}), (\ref{CH}) and (\ref{AH}), which can be nicely
organized in the  following table:

\begin{equation}
\begin{array}{ll}
dS~momentum-space\quad &\quad dS~spacetime \nonumber\\
\,&\,\\
{\cal H}_\ell=p_1-\frac{1-e^{-\ell p_0}}{\ell} & {\cal H}_H=p_1+p_0 e^{H x^0}, \nonumber\\
\,&\,\\
{\cal A}_\ell=x^1+x^0 e^{\ell p_0} & {\cal A}_H=x^1-\frac{1-e^{-H x^0}}{H}\nonumber,
\end{array}
\end{equation}

\section{CONCLUSIONS}
We feel we here provided a satisfactory understanding of the travel-time features and of some aspects of relative
spacetime locality which had been encountered in previous studies of theories with curved momentum space,
but for which a clear conceptual picture was still missing.

We expect it should be possible to use the results here reported as starting point for a similar understanding
of other features and other manifestations of relative spacetime locality encountered in studies
of theories with curved momentum spaces.
Of particular interest from this perspective could be the results reported in Refs.\cite{leelaurentGRB,transverse}
on dual-gravity lensing and transverse relative locality. These arise from momentum spaces which do not have de Sitter geometry and produce novel effects and novel implications for spacetime locality along directions
orthogonal to the one connecting the emitter and the detector (they do not affect the travel times but rather
the directional information codified in the description of the relevant processes).

Concerning relative spacetime locality and its connection with manifestations of ordinary curved-spacetime-induced
as relativity of momentum-space locality our analysis here was facilitated by the fact that we have dealt
exclusively with the idealized case of non-interacting particles. One can look back at our findings and
abstract the observation that relative locality is a feature arising whenever the chosen coordinatization
does not have simple properties  under the relativistic-symmetry transformation of interest.
In particular, for the translation transformations here considered relative locality arises when
the coordinates have non-canonical Poisson brackets with the generators of translations.
It is noteworthy that in the case of non-interacting particles one could always choose coordinates which are
free from the relative-locality effects. Such coordinates are not always the most convenient
(depending on what are the objectives on one's study) but they are always available in theories
with only non-interacting particles. It appears that this aspect of simplicity should be lost
for interacting particles in presence of curvature, or at least this is what is suggested by
studies~\cite{leelaurentGRB,anatomy} of the recently proposed ``relative-locality framework"~\cite{principle}. That framework allows to describe interactions
compatibly with the presence of curvature on momentum space, but then the translation generators
acquire a novel form such that one cannot provide for each (interacting) particle in the system
a coordinatization with canonical Poisson brackets with the generators of translation transformations.
It would be interesting to find an analogue of this feature of interacting theories with a curved momentum
space in some theories with a curved spacetime.

\bigskip

$~$

\bigskip

$~$

\noindent
{\it
This research was supported in part by the John Templeton Foundation.}

\end{document}